\documentclass{article} 
\usepackage{iclr2020_conference,times}
\usepackage{graphicx}
\iclrfinalcopy

\usepackage{hyperref}
\usepackage{url}

\title{The DIDI dataset: Digital Ink Diagram data}

\author{Philippe Gervais \& Thomas Deselaers\\Google Research\\\texttt{\{pgervais, deselaers\}@google.com} \And Emre Aksan \& Otmar Hilliges\\ETH Zurich\\\texttt{\{eaksan,otmar.hilliges\}@inf.ethz.ch}}

\begin{document}

\maketitle

\begin{abstract}
We are releasing a dataset of diagram drawings with dynamic drawing information. The dataset aims to foster research in interactive graphical symbolic understanding. The dataset was obtained using a prompted data collection effort.
\end{abstract}

\section{Introduction}
\label{sec:intro}
Handwritten text and hand-drawn sketches have been one of our primary means of processing, conveying and storing information. While optical character recognition (OCR) is one of the early approaches converting handwritten content into digital representation, digital ink allows for combining the flexibility and aesthetics of handwriting with the ability to process and edit interactively.

A vast range of mobile computing devices with touch screens provide interaction by means of digital ink in different forms. Smart watches or mobile phones enable users to write and draw on the screen. High-end laptops and tablets \citep{ipadpro,surface,pixelbook}, often with a stylus, allow to do fine-grained note-taking and document editing. Ink-based applications provide natural interaction and richer user experience. Handwriting recognition is a a natural text input method \citep{yaeger:aaai98,pittman:2007,google-hwrlstm}. Recent works focus on assisting tools to preserve and enhance stylistic aspects of handwriting \citep{zitnick2013handwriting} as well as content editing \citep{aksan2018deepwriting}. \cite{ha2017neural} present a predictive model for completing hand-drawn objects.

While existing work has been successful on modelling and recognition of isolated hand-written and -drawn content, we argue that the real potential of digital ink lies in modeling of richer content where multiple categories such as handwritten text, hand-drawn shapes, and diagrams co-exist and are semantically related. To foster the research into this direction, we release the novel \emph{DIDI} dataset of digital ink diagrams. DIDI consists of large number of user-drawn diagram samples with and without text, allowing for data-driven modelling. 

We hope to create more interest in the academic community with the release of this dataset that enables research into the combination of machine learning and human computer interaction - which we believe both to be important for a practical diagram creation and editing tool. 


\section{Related Work}
\label{sec:related}

To the best of our knowledge only few datasets of handwritten drawings are available to date: 

\begin{enumerate}
\item
Czech Technical University has released a dataset of diagram drawings with a total of 300 finite automata diagrams and 672 flowchart diagrams \citep{bresler:icfhr2014,bresler:ijdar2016}
\item
The Nakagawa lab at Tokyo University of Agriculture and Technology has released the KONDATE dataset of 670 drawings \citep{kondate}.
\item
University of Nantes has released a dataset of 419 flowchart drawings collected from 36 writers \citep{awal:hal-00518451}.
\end{enumerate}

Another related field of more specialized drawings is the creation of mathematical equations, another domain where using a freeform input maybe a big advantage in how natural and efficiently users are enabled to create digital content. In this field, the CROHME competition \citep{chrohme-ijdar} has created a dataset of more than 10,000 mathematical expressions, including the dynamic time information since 2011. 

Unfortunately, none of these datasets is large enough to be considered large from a deep learning perspective. 
\section{Dataset Description}
\label{sec:dataset}

We are releasing a dataset of flowchart drawings consisting of two parts: 22,287 diagrams with textual labels and 36,368 diagrams without textual labels. The data was collected from a total of 364 participants where the number of individual drawings any participant created was between 1 and 1291.

The data was collected in a prompted data collection effort where the participants were shown an image and asked to draw the shown diagram over the image (see figure 1).

We are releasing the data online under the Creative Commons Attribution 4.0 International License at \url{https://github.com/google-research/google-research/tree/master/didi_dataset} with example code demonstrating how to visualize the data and how to convert it into formats suitable to common machine learning frameworks.

\begin{figure}[tbp]
\begin{center}

\includegraphics[width=.3\linewidth]{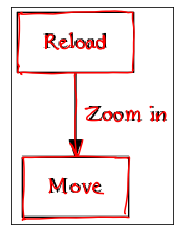}
\includegraphics[width=.3\linewidth]{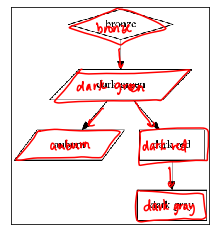}
\includegraphics[width=.3\linewidth]{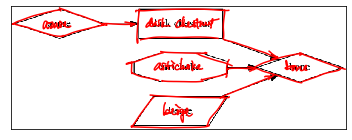}
\includegraphics[width=.3\linewidth]{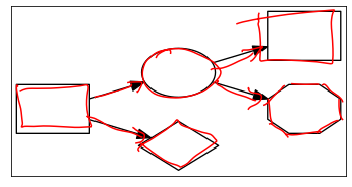}
\includegraphics[width=.3\linewidth]{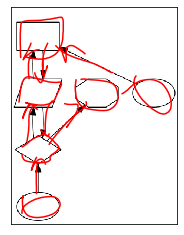}
\includegraphics[width=.3\linewidth]{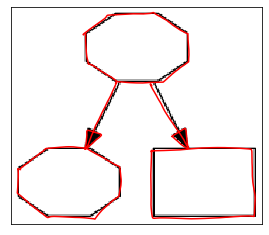}

\caption{Example diagrams with (first row) and without textual content (second row) overlayed with their respective prompt images.}
\label{fig:example_image}
\end{center}
\end{figure}

In figure~\ref{fig:example_image} we show a number of example drawings overlaid with the respective prompt images.

In the following we describe the data collection setup, the data format, and the dataset statistics.

\subsection{Data collection}
\label{sec:datacollection}
The data collection was performed using an Android app on Chromebooks (with and without stylus) and Android devices. The app is used for a variety of data collection tasks related to online handwriting recognition. Participants were mostly interns at Google's Zurich (Switzerland) and Mountain View (California, US) offices.

\begin{figure}[tb]
\begin{center}
\includegraphics[width=.5\linewidth]{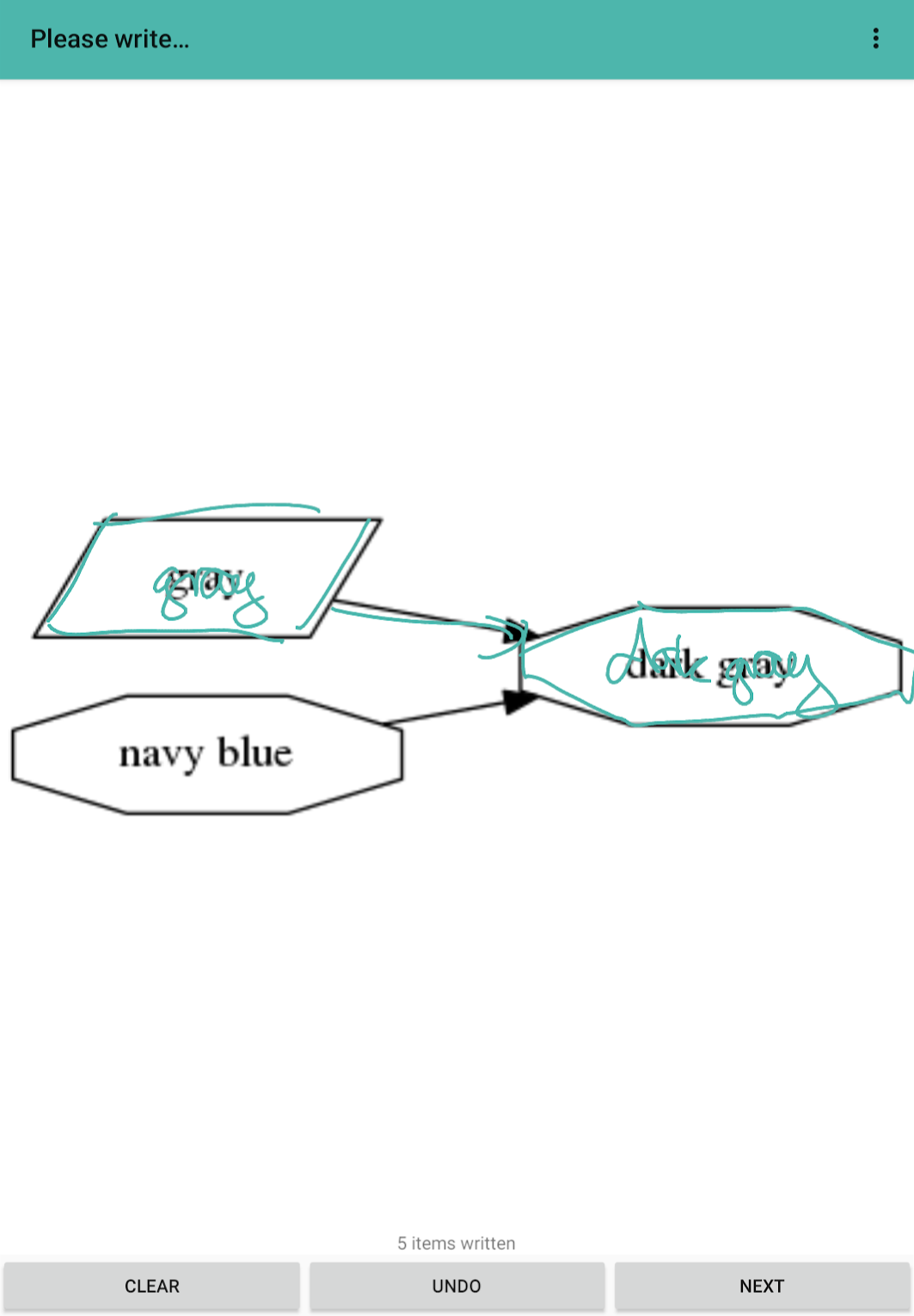}
\caption{A screenshot of the data collection app. The user has already partially drawn (in green) the prompt (in black). The app allows to start from scratch (clear button), remove the last stroke (undo), or proceed to the next drawing, which stores the users' drawing.}
\label{fig:collectionapp}
\end{center}
\end{figure}

The app was configured to show a prompt image of a flowchart to the user who was then asked to draw over it. See figure \ref{fig:collectionapp} for an example. 

Flowcharts images were obtained through GraphViz, based on randomly-generated dot files. See section~\ref{sec:generation} for a description on how the prompt images were generated. 

The data collection was organized in two sessions. The first session happened in summer 2018 and participants were asked to draw diagrams like the one shown in figure~\ref{fig:collectionapp}, including the textual content in the boxes. The second session happened in summer 2019 and participants were asked to draw the same type of diagrams but without any textual content in the boxes.


\subsection{Data format}
\label{sec:dataformat}
The data is released in NDJSON format (newline delimited JSON) with additional supporting files. Each row of the files corresponds to one drawing and contains a number of fields:

\begin{description}
\item[key] a 64-bit integer as an hexadecimal string - a unique identifier for that row.
\item[label id] a sha1 hash of the dot file that was used to generate the prompt image. Labels occur between 1 and 169 times in the dataset. For each label we also provide the original dot file that was used to render the prompt image, the resulting PNG image, and the extended xdot file generated by GraphViz. 
\item[drawing] the drawing itself as an array of strokes, where each stroke is an array containing arrays for x, y, coordinates and timestamps (in milliseconds, starting at zero for each drawing). 
\item[writing guide] the width and height of the drawing area where the user has drawn. This information can be used to align the prompt image with the coordinates. See the accompanying colab notebook for more detail.
\item[split] one of \texttt{train}, \texttt{valid}, or \texttt{test}, indicating whether 
\end{description}

An example row from one of the NDJSON files is shown in figure~\ref{fig:ndjson}.

\begin{figure}[tbp]
\begin{verbatim}
{"key": "00045b63775fd0ca", 
 "label_id": "5a663549e8a68735ede3f7ef38542fda4fd7693e", 
 "split": "test",
 "writing_guide": {"width": 1066, "height": 1256}, 
 "drawing": [[[21.7, 19.3,...], [610.5, 607.3, ...], [0, 12, ...]], 
             [[...], [...], [...]],
             ,...
            ]
 }
\end{verbatim}
\caption{A (shortened) line from the NDJSON file.}
\label{fig:ndjson}
\end{figure}

\subsection{Diagram Generation Process}
\label{sec:generation}

Diagrams used as prompts were generated using GraphViz, from randomly generated dot files. The dotfiles were generated as follows:

\begin{enumerate} 
\item Select a number of nodes between 2 and 6 uniformly
\item Select a text label topic from \{none, colors,  process, clients\}
\item graph\_has\_labeled\_edges := Bernoulli(p=.7)
\item For each node
\begin{enumerate}
\item randomly assign a shape from \{box, oval, diamond, parallelogram, octagon\}
\item randomly assign a textual label from the topic selected above
\end{enumerate}
\item while \#edges $<$ \#nodes - 1 or there are unconnected nodes
\begin{enumerate}
 \item while true:
 \begin{enumerate}
   \item randomly select a start and stop node
   \item if no edge already exists between the nodes, create it and break from the loop
 \end{enumerate}
 \item if graph\_has\_labeled\_edges: randomly assign a label to the new edge
\end{enumerate}
\end{enumerate}

The possible labels for each of the topic are given in Appendix~\ref{app:wordlists}.

\subsection{Dataset statistics}
\label{sec:datasetstats}

The complete dataset of 58655 drawings was collected from a total of 364 participants. 
The dataset uses a total of 6555 unique labels, with each label being used between 1 and 169 times, median 4. 

The collected drawings use between 1 and 161 strokes (median 14) and contain between 2 and 17980 points (median 1559).

\subsection{Experimental protocol}
\label{sec:splits}

For uses of the dataset where dedicated training, validation, and test sets are desirable, we define a split in the NDJSON files. 
The split was created based on writers (whose identifiers are not released) such that the sets of writers in the respective training, validation, and test parts of the datasets are disjoint.

Given the size of the dataset, we chose to use about 1/8th of the data for validation and test data, respectively, and to use the remaining 6/8th of the data as training set. 

\begin{table}[tb]
    \caption{Sizes of the training/validation/test parts of the dataset.}
    \label{tab:splits}
    \centering
    \begin{tabular}{|l|r|r|r|}
    \hline
         Split  & w/o text & w/ text & all \\
    \hline
         Training &   27,278  & 16,717  & 43,995   \\ 
         Valid    &    4,545  &  2,785  &  7,330   \\ 
         Test     &    4,545  &  2,785  &  7,330   \\
   \hline
    \end{tabular}
\end{table}

\section{Conclusion}
\label{sec:conclusion}
We are releasing a dataset of online diagram drawings with their respective prompt in order to foster research in graphical symbolic understanding by combining machine learning and human computer interaction technologies.

\bibliography{diagrams.bib}

\begin{thebibliography}{14}
\providecommand{\natexlab}[1]{#1}
\providecommand{\url}[1]{\texttt{#1}}
\expandafter\ifx\csname urlstyle\endcsname\relax
  \providecommand{\doi}[1]{doi: #1}\else
  \providecommand{\doi}{doi: \begingroup \urlstyle{rm}\Url}\fi

\bibitem[Aksan et~al.(2018)Aksan, Pece, and Hilliges]{aksan2018deepwriting}
Emre Aksan, Fabrizio Pece, and Otmar Hilliges.
\newblock Deepwriting: Making digital ink editable via deep generative
  modeling.
\newblock In \emph{Proceedings of the 2018 CHI Conference on Human Factors in
  Computing Systems}, pp.\  1--14, 2018.

\bibitem[Apple()]{ipadpro}
Apple.
\newblock {iPad Pro}.
\newblock \url{https://www.apple.com/ipad-pro/}.
\newblock Accessed: 2020-02-14.

\bibitem[Awal et~al.(2011)Awal, Feng, Mouchère, and
  Viard-Gaudin]{awal:hal-00518451}
Ahmad-Montaser Awal, Guihuan Feng, Harold Mouchère, and Christian
  Viard-Gaudin.
\newblock First experiments on a new online handwritten flowchart database.
\newblock In \emph{Document Recognition and Retrieval XVIII}, 2011.

\bibitem[Bresler et~al.(2014)Bresler, Phan, Průša, Nakagawa, and
  Hlaváč]{bresler:icfhr2014}
Martin Bresler, Truyen~Van Phan, Daniel Průša, Masaki Nakagawa, and Václav
  Hlaváč.
\newblock Recognition system for on-line sketched diagrams.
\newblock In \emph{ICFHR}, 2014.

\bibitem[Bresler et~al.(2016)Bresler, Průša, and Hlaváč]{bresler:ijdar2016}
Martin Bresler, Daniel Průša, and Václav Hlaváč.
\newblock Online recognition of sketched arrow-connected diagrams.
\newblock \emph{IJDAR}, 2016.

\bibitem[Carbune et~al.(2020)Carbune, Gonnet, Deselaers, Rowley, Daryin, Calvo,
  Wang, Keysers, Feuz, and Gervais]{google-hwrlstm}
Victor Carbune, Pedro Gonnet, Thomas Deselaers, Henry~A. Rowley, Alexander
  Daryin, Marcos Calvo, Li-Lun Wang, Daniel Keysers, Sandro Feuz, and Philippe
  Gervais.
\newblock Fast multi-language {LSTM}-based online handwriting recognition.
\newblock \emph{IJDAR}, 2020.

\bibitem[Google()]{pixelbook}
Google.
\newblock {PixelBook}.
\newblock \url{https://store.google.com/product/google_pixelbook}.
\newblock Accessed: 2020-02-14.

\bibitem[Ha \& Eck(2017)Ha and Eck]{ha2017neural}
David Ha and Douglas Eck.
\newblock A neural representation of sketch drawings.
\newblock \emph{arXiv preprint arXiv:1704.03477}, 2017.

\bibitem[Matsushita \& Nakagawa(2014)Matsushita and Nakagawa]{kondate}
Tomohisa Matsushita and Masaki Nakagawa.
\newblock A database of on-line handwritten mixed objects named 'kondate'.
\newblock In \emph{ICFHR}, 2014.

\bibitem[Microsoft()]{surface}
Microsoft.
\newblock {Microsoft Surface}.
\newblock \url{https://www.microsoft.com/en-us/surface/}.
\newblock Accessed: 2020-02-14.

\bibitem[Mouchère et~al.(2016)Mouchère, Zanibbi, Garain, and
  Viard-Gaudin]{chrohme-ijdar}
Harold Mouchère, Richard Zanibbi, Utpal Garain, and Christian Viard-Gaudin.
\newblock Advancing the state of the art for handwritten math recognition: the
  crohme competitions, 2011–2014.
\newblock \emph{IJDAR}, 2016.

\bibitem[Pittman(2007)]{pittman:2007}
James~A. Pittman.
\newblock Handwriting recognition: Tablet {PC} text input.
\newblock \emph{{IEEE} Computer}, 40\penalty0 (9):\penalty0 49--54, 2007.

\bibitem[Yaeger et~al.(1998)Yaeger, Webb, and Lyon]{yaeger:aaai98}
Larry Yaeger, Brandyn Webb, and Richard Lyon.
\newblock Combining neural networks and context-driven search for on-line,
  printed handwriting recognition in the {N}ewton.
\newblock \emph{AAAI AI Magazine}, 1998.

\bibitem[Zitnick(2013)]{zitnick2013handwriting}
C~Lawrence Zitnick.
\newblock Handwriting beautification using token means.
\newblock \emph{ACM Transactions on Graphics (TOG)}, 32\penalty0 (4):\penalty0
  1--8, 2013.

\end{thebibliography}
\bibliographystyle{iclr2020_conference}

\newpage
\appendix
\section{Appendix: Statistics}
\begin{itemize}
\item Total number of diagrams:
\begin{itemize}
\item with text: 22287
\item without text: 36368
\end{itemize}
\item Total number of nodes: 
\begin{itemize}
\item within graphs with text: 77682
\item within graphs without text: 139953
\end{itemize}

\item Total number of edges
\begin{itemize}
\item with a text label: 17865
\item within graphs with text: 64361
\item within graphs without text: 122535
\end{itemize}
\item Node shapes:
\begin{itemize}
\item within graphs with text: 
\begin{itemize}
\item box 15906
\item diamond 15360
\item octagon 15620
\item oval 15384
\item parallelogram 15412
\end{itemize}

\item within graphs without text:
\begin{itemize}
\item box 28251
\item diamond 27886
\item octagon 29115
\item oval 27575
\item parallelogram 27126
\end{itemize}
\end{itemize}
\end{itemize}

\newpage
\section{Appendix: Word lists}
\label{app:wordlists}

The word lists used for each of the topics are given in table \ref{tab:wordlists}.

\begin{table}[h]
\caption{Words used as node and edge labels in the generation of the diagrams (cp.\ section~\ref{sec:generation}). The numbers in parentheses indicate how often each label occurs in the dataset.}
\begin{center}
\begin{tabular}{|l|p{5.5cm}|p{5.5cm}|}
\hline
Topic & Node labels & Edge labels\\
\hline
None & -- & -- \\\hline
Colors &
artichoke (680),
auburn (684),
azure (732),
black (699),
blue (746),
bronze (654),
brown (650),
bubble gum (611),
charcoal (742),
chestnut (725),
chocolate (707),
cobalt blue (627),
cyan (702),
dark brown (1389),
dark chestnut (575),
dark chocolate (652),
dark cyan (669),
dark gray (617),
dark green (729),
dark red (640),
eggshell (680),
fuchsia (652),
gold (738),
gray (663),
green (734),
help (412),
light blue (634),
light brown (726),
light green (661),
lime (785),
magenta (737),
navy blue (736),
orange (576),
pink (727),
purple (632),
red (756),
rose (648),
ruby (770),
saffron (720),
silver (608),
teal (640),
violet (679),
white (680),
yellow (730)
& --
\\\hline

Process&
Wait (1014),
Plug (1113),
Resume (928),
Move (1066),
Go (1035),
Turn on (1047),
Quit (936),
Create (1066),
Ready? (1099),
Search (1040),
Reload (997),
Delete (992),
Replace (959),
Quit? (981),
Switch (1186),
Repair (1110),
Close (1029),
Open (1070),
Turn off (969),
Unplug (1128),
Stop (1102),
Copy (998),
Advance (966)
&
bills (720),
builds (692),
calls (700),
connects (677),
drops (684),
helps (675),
manufactures (649),
processes (774),
produces (694),
replies (693),
sends (746),
ships (665)
\\\hline

Clients &
account no (435),
address (484),
bank (420),
billing (449),
business (395),
city (485),
communication (366),
customer (456),
customer care (337),
customs (433),
documentation (469),
email (470),
employee (346),
evaluation (448),
feedback (448),
fees (406),
first name (415),
hotline (436),
id (434),
last name (398),
letter (405),
location (376),
manager (428),
market (425),
name (355),
number (450),
order (462),
payment (449),
payroll (412),
phone (527),
place (417),
price (433),
process (417),
product (444),
promotion (485),
purchase (446),
recipient (356),
sales (477),
sender (452),
shipment (508),
street (457),
street number (407),
support (468),
target (444),
taxes (508),
website (426),
zip code (466)
&
bills (720),
builds (692),
calls (700),
connects (677),
drops (684),
helps (675),
manufactures (649),
processes (774),
produces (694),
replies (693),
sends (746),
ships (665)
\\\hline

\end{tabular}
\end{center}
\label{tab:wordlists}
\end{table}%

\end{document}